%% file: iclr2026_Main.tex
\documentclass{article}

\PassOptionsToPackage{numbers,compress}{natbib}
\usepackage[preprint]{neurips_2025}

\usepackage[utf8]{inputenc}
\usepackage[T1]{fontenc}
\usepackage{hyperref}
\usepackage{url}
\usepackage{booktabs}
\usepackage{amsmath,amsfonts,amssymb}
\usepackage{microtype}
\usepackage{xcolor}
\usepackage{graphicx}
\usepackage{subcaption}
\usepackage{multirow}
\usepackage{array}
\usepackage{enumitem}
\usepackage{nicefrac}
\usepackage{placeins}

\graphicspath{{Figs/}}
\setlist[itemize]{leftmargin=*,topsep=2pt,itemsep=2pt}

\title{A Low-Power Wearable Respiratory Sensor for Non-Invasive Stress Monitoring}

\author{%
  Hamed Khatounabadi\textsuperscript{\textdagger}\\
  Department of Electrical Engineering\\
  Sharif University of Technology\\
  Tehran, Iran\\
  \And
  Mohammad Hosseini\textsuperscript{\textdagger} \\
  Department of Electrical Engineering\\
  Sharif University of Technology\\
  Tehran, Iran\\
  \AND
  Mohammad Fakharzadeh\\
  Department of Electrical Engineering\\
  Sharif University of Technology\\
  Tehran, Iran\\
}

\begin{document}

\maketitle
\begingroup
\renewcommand{\thefootnote}{\fnsymbol{footnote}}
\footnotetext[2]{Equal contribution.}
\endgroup

\begin{abstract}
\input{abstract}
\end{abstract}

\input{introduction}
\FloatBarrier
\input{methods}
\FloatBarrier
\input{results}
\input{discussion}
\FloatBarrier

\bibliographystyle{plainnat}
\bibliography{references}

\FloatBarrier
\appendix
\input{appendix}

\end{document}

%% file: abstract.tex
Respiration provides a continuously available window into physiological state and behavior. However, monitoring it outside controlled settings remains challenging because a wearable system must capture small body deformations while remaining comfortable, low power, and robust to changes in posture and motion. We present a compact non-invasive respiratory sensing system based on a force-sensitive resistor (FSR) embedded in an abdominal belt and integrated with a custom Bluetooth Low Energy acquisition board. The system combines a simple piezoresistive readout with a mechanical holder designed to transfer abdominal expansion to the sensor without analog amplification. We evaluate the complete sensing pipeline across multiple breathing patterns and body positions. In stationary settings, the recorded signals exhibit consistent amplitude changes and recurring peak-to-peak timing across breathing maneuvers; under light movement, these variations remain visible despite motion-induced baseline shifts. We further design a five-phase stress-induction protocol and collect respiratory recordings from 12 participants. Using interpretable time-domain features and standard classifiers, we examine whether the acquired signals distinguish relaxation from stress-induction phases. In this preliminary experiment, the best-performing model achieves 88.0\% test accuracy, indicating that the extracted respiratory features distinguish stress-induced phases from relaxation phases in this dataset. Overall, our results show that the proposed platform enables real-time respiratory monitoring across diverse daily-life scenarios and captures respiratory changes that distinguish stress-induction from relaxation phases, supporting its potential for affective-computing applications.

%% file: introduction.tex
\section{Introduction}

Respiration is one of the most accessible physiological signals for monitoring health and behavior. Its rate and waveform reflect both metabolic demand and autonomic regulation, and changes in respiration have been associated with clinical deterioration, exercise, sleep and pulmonary disorders, pain, cognitive load, stress, and anxiety \citep{fieselmann1993respiratory,cretikos2008respiratory,nicolo2020importance,grassmann2016respiratory,homma2008breathing}. Unlike many measurements that are restricted to clinical or laboratory settings, respiration can in principle be recorded continuously using small wearable devices \citep{chu2019respiration,chen2021individuality}. This makes respiratory monitoring promising for longitudinal health assessment and affective computing \citep{hovsepian2015cstress,schmidt2018wesad}.

Despite this promise, reliable continuous measurement remains difficult outside controlled settings. Reference instruments such as spirometers and respiratory inductive plethysmography are effective in supervised experiments, while impedance, airflow, acoustic, optical, thermal, inertial, and deformation-based methods each introduce different compromises in comfort, accuracy, power consumption, and sensitivity to environmental or motion artifacts \citep{folke2003critical,alkhalidi2011respiration,massaroni2019contact,charlton2018breathing}. A practical wearable must resolve the comparatively small deformations caused by breathing while maintaining stable mechanical contact as posture, placement, and belt tension change. It must also support battery-powered acquisition and wireless transmission without making the device cumbersome \citep{hussain2023wearable,yin2024wearable}. Satisfying these requirements within one compact system remains an important engineering challenge.

Obtaining a reliable respiratory waveform is only the first part of the problem. For downstream classification of stress or other labeled physiological states, the recorded time series must be transformed into features or learned representations that preserve informative temporal patterns while reducing noise and nuisance variability. This feature-extraction step is a central challenge in physiological time-series processing, especially when labeled data are limited and class-related changes may be subtle. Recent work on dynamical modeling and temporal representation learning illustrates how structured models can extract informative representations from complex time series \citep{hosseini2025dynamical,hosseini2026crosssubject,sani2026preferential}. These considerations motivate treating respiratory acquisition, temporal feature extraction, and downstream classification as parts of one coupled processing pipeline.

\paragraph{Contributions.}
Here, we present a low-cost wearable respiratory-sensing system that combines a force-sensitive resistor (FSR), a mechanically compliant abdominal holder, a rechargeable power supply, and a compact Bluetooth Low Energy (BLE) acquisition board. Abdominal expansion changes the pressure applied to the piezoresistive element, a voltage divider converts this resistance change into an analog signal, and the custom board digitizes and transmits the resulting waveform. The design deliberately favors a mechanically effective pressure-transfer structure and a large raw signal swing over a more elaborate analog front end.

Our contributions are threefold: \textbf{1)} we develop the complete wearable device, including its passive FSR readout, abdomen-mounted holder, rechargeable supply, and custom BLE PCB; \textbf{2)} we collect a prototype-scale respiratory dataset and show that the device produces usable signals with distinct amplitude and timing behavior across normal, deep, fast, and breath-hold patterns and across sitting, lying, standing, and walking positions; and \textbf{3)} we design a five-phase stress-induction protocol, collect respiratory recordings from 12 participants, and demonstrate the potential of the acquired respiratory signals for affective computing through preliminary binary classification of relaxation and stress-induction phases, with the best model reaching 88.0\% test accuracy under the original segment-level evaluation.

\section{Related Work}

\paragraph{Respiratory monitoring modalities.}
Respiration can be measured directly from airflow or gas exchange, indirectly through respiratory modulation of other physiological signals, or mechanically from thoracic and abdominal motion \citep{alkhalidi2011respiration,folke2003critical,massaroni2019contact,hussain2023wearable}. ECG- and photoplethysmography-derived approaches can reuse existing physiological channels, but their accuracy depends on the strength of respiratory modulation and on the estimation algorithm \citep{charlton2018breathing}. Contactless camera, thermal, ultrasound, and radar systems avoid body attachment, but their performance depends on viewing geometry, environmental conditions, and the separation of breathing from other body movements \citep{alkhalidi2011respiration,machado2018}. In contrast, mechanical wearables measure local changes in circumference, strain, pressure, or acceleration and can provide a direct, low-latency respiratory waveform \citep{massaroni2019contact,hussain2023wearable}. For applications that require continuous monitoring, these tradeoffs motivate compact wearable sensors that can acquire respiration directly on the body.

\paragraph{Wearable deformation sensors.}
Mechanical respiratory wearables have been developed using inductive bands, capacitive structures, piezoelectric elements, strain gauges, piezoresistive textiles, elastomer sensors, and optical fibers \citep{massaroni2019contact,hussain2023wearable,yin2024wearable}. Among these approaches, piezoresistive devices are particularly compatible with compact electronics because their resistance changes can be measured using simple readout circuits. Prior work has demonstrated 3D-printed piezoresistive enclosures, disposable strain patches, garment-integrated resistive sensors, and wireless chest or abdominal devices \citep{vanegas2019piezoresistive,chu2019respiration,kim2024accurate}. However, these studies also show that the measured waveform depends strongly on calibration, placement, and the mechanical coupling between the body and sensor. Our system builds on this class of sensors while emphasizing a low-cost implementation and the combined design of the FSR holder, embedded electronics, and acquisition pipeline.

\paragraph{Motion robustness and pattern modeling.}
Motion and placement remain central challenges for deformation-based respiratory sensing because gross body movement can exceed respiration-induced deformation. Walking, posture changes, garment slip, and variations in attachment pressure can distort the waveform or shift the sensor operating point. Prior work has demonstrated respiration sensing under ambulatory conditions and accelerometer-referenced adaptive filtering for reducing motion artifacts \citep{chu2019respiration,zhang2014adaptive}. Related studies address the complementary problem of preserving useful respiratory representations across posture and breathing-pattern variability: wireless chest and abdominal sensors characterize posture-dependent and individual respiratory behavior \citep{chen2021individuality}; inertial wearables with convolutional models distinguish simulated breathing events \citep{mcclure2020classification}; and multisensor flex and inertial systems use deep sequence models for breathing-pattern recognition \citep{comeau2026aienabled}. Our results likewise show that the single-FSR pipeline preserves distinguishable respiratory structure across the tested sitting, lying, standing, and walking conditions, demonstrating robustness across a range of posture and light-motion scenarios.

\paragraph{Respiration and stress inference.}
Beyond respiratory-rate estimation, waveform timing, amplitude, and pattern changes can provide information about breathing behavior and physiological state \citep{chen2021individuality,mcclure2020classification,comeau2026aienabled}. Respiration changes with emotion and cognitive demand through both autonomic regulation and task-dependent behaviors such as speaking, sighing, and breath holding \citep{boiten1998effects,homma2008breathing,grassmann2016respiratory}. The Stroop Color--Word Task and mental arithmetic are established laboratory stressors, and prior studies have reported changes in respiration and other psychophysiological measurements during these tasks \citep{stroop1935studies,tulen1989characterization,cipresso2019computational}. Automated stress-detection systems commonly combine respiration with ECG, heart-rate variability, electrodermal activity, temperature, electromyography, or motion \citep{healey2005detecting,sun2010activity,hovsepian2015cstress,schmidt2018wesad,gjoreski2016continuous}, while respiration-only approaches have also been explored using thermal imaging and bioradar \citep{cho2017deepbreath,machado2018}. Our preliminary experiment builds on this literature and shows that features extracted from the acquired respiratory signals distinguish relaxation and stress-induction phases, demonstrating their potential for downstream affective-computing applications.

%% file: methods.tex
\section{Methods}

\subsection{Sensor Selection and Sensing Mechanism}
\label{sec:sensor_selection}

A key component of the proposed system is the selection of an appropriate sensing modality for capturing respiratory signals. Among accelerometers, capacitive sensors, piezoelectric sensors, strain gauges, and piezoresistive sensors, the selection process was guided by three criteria: measurement accuracy, system complexity, and cost efficiency \citep{massaroni2019contact,hussain2023wearable}.

Accelerometer-based and capacitive sensors were initially considered; however, they were less suitable for the target wearable setting. Inertial measurements are strongly affected by non-respiratory body motion, and capacitive sensing can require careful calibration and shielding because body proximity, skin properties, and sensor geometry all affect the measured capacitance \citep{massaroni2019contact,zhang2014adaptive}. We therefore focused on piezoelectric, strain-gauge, and piezoresistive sensing technologies for the final design.

Piezoelectric sensors generate electrical charge in response to applied force and are useful for dynamic, high-frequency mechanical events. They are less suitable for static or slowly varying pressure because the generated charge can leak over time, which is problematic during apnea or breath-hold intervals. Strain gauges measure resistance changes caused by mechanical deformation and can provide accurate measurements, but they usually require a Wheatstone bridge, amplification circuitry, and careful noise control. Many low-cost strain gauges also have low nominal resistance, which increases current consumption and makes them less attractive for long-term battery-powered operation.

Based on this analysis, we selected a force-sensitive resistor (FSR) as the primary sensing element. The FSR is a passive piezoresistive sensor whose resistance decreases as applied pressure increases. It provides a practical trade-off between simplicity, flexibility, low cost, low power consumption, and adequate sensitivity. In the unloaded state, its resistance can exceed the megaohm range and then decreases by several orders of magnitude as the applied force increases \citep{interlinkFSR402}. This allows the sensor to be read with a simple voltage divider rather than a specialized analog front end, consistent with the low-complexity readout used in related piezoresistive respiratory wearables \citep{vanegas2019piezoresistive,chu2019respiration,kim2024accurate}.

The sensing circuit is integrated into this first methodological stage because it is part of the sensor-selection rationale. In the implemented divider, the FSR is connected between the divider supply, $V_{\mathrm{dd}}$, and the ADC node, while the fixed load resistor, $R_L=10~\mathrm{k}\Omega$, is connected between that node and ground. The initial Arduino prototype used a $5~\mathrm{V}$ divider supply, whereas the custom PCB uses its regulated $1.8~\mathrm{V}$ rail. The $10~\mathrm{k}\Omega$ value is an implementation choice that provides a useful output swing over the observed FSR range:
\begin{equation}
    V_{\mathrm{out}} = V_{\mathrm{dd}}\frac{R_L}{R_{\mathrm{FSR}} + R_L}.
    \label{eq:divider}
\end{equation}
Thus, inhalation-induced abdominal expansion increases the applied pressure, reduces $R_{\mathrm{FSR}}$, and increases the voltage measured at the ADC input.

\subsection{Prototype Design and Wearable Setup}
\label{sec:prototype_design}

\begin{figure}[t]
    \centering
    \begin{subfigure}{0.48\linewidth}
        \centering
        \includegraphics[width=\linewidth]{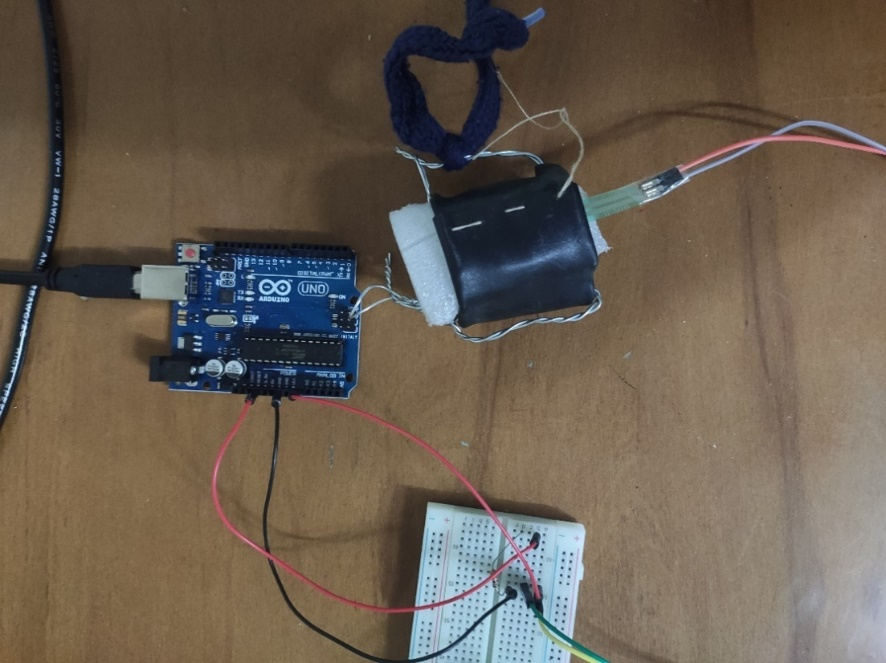}
        \caption{Prototype circuit and materials}
    \end{subfigure}
    \hfill
    \begin{subfigure}{0.48\linewidth}
        \centering
        \includegraphics[width=\linewidth]{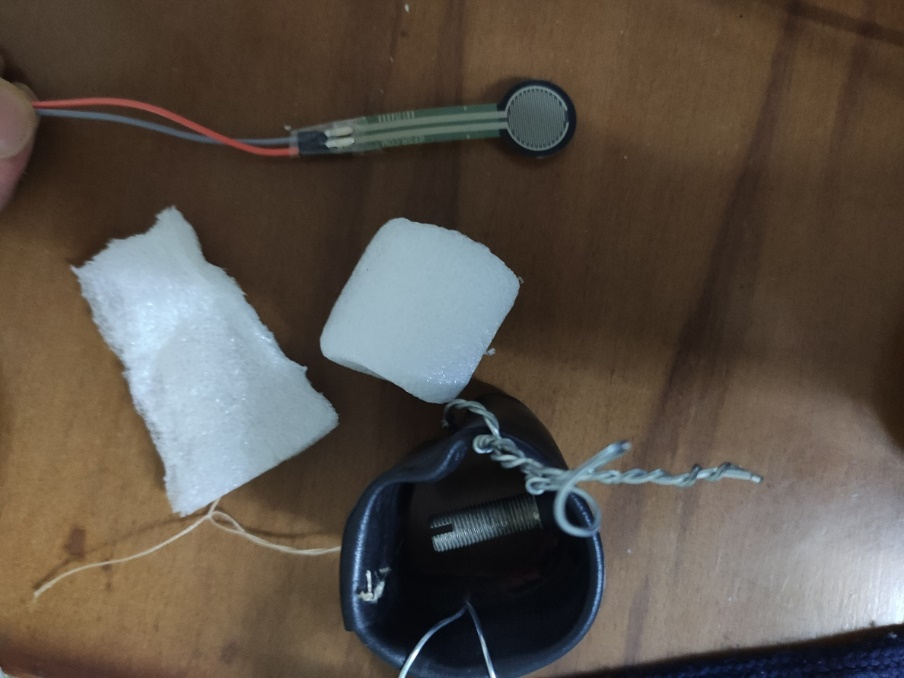}
        \caption{Layered wearable materials}
    \end{subfigure}
    \caption{Initial prototype used for validating the FSR sensing concept before custom PCB fabrication.}
    \label{fig:prototype_overview}
\end{figure}

To validate the feasibility of the sensing approach before designing a dedicated printed circuit board, we first developed a wearable prototype using readily available components and materials. The goal of this stage was to test whether the FSR could reliably capture respiratory pressure variations under realistic body placement conditions. Figure~\ref{fig:prototype_overview} shows this first feasibility prototype.

\subsubsection{Prototype Construction}

For signal sampling and data acquisition, an Arduino UNO in the DIP package was used, and the divider output was connected directly to its ADC input. The initial prototype used soft and elastic materials to stabilize the sensor while preserving user comfort. A bicycle inner tube was selected because of its elasticity. The valve section was cut, and the two ends were stitched together to form a closed loop. Since the metal valve could concentrate pressure and damage the FSR, it was covered with protective foam. The inner contact surface between the tube and sensor was also padded with foam to distribute pressure more uniformly.

Both ends of the tube were reinforced using wire loops that acted like belt-buckle anchors. An elastic strap was attached to these loops so the structure could be worn around the abdomen. This arrangement converted abdominal expansion and contraction (Figure~\ref{fig:respiration_mechanism}) into a localized pressure change on the FSR, while avoiding direct rigid contact between the sensor and body.

\subsubsection{Respiratory Signal Acquisition Mechanism}

\begin{figure}[t]
    \centering
    \includegraphics[width=0.82\linewidth]{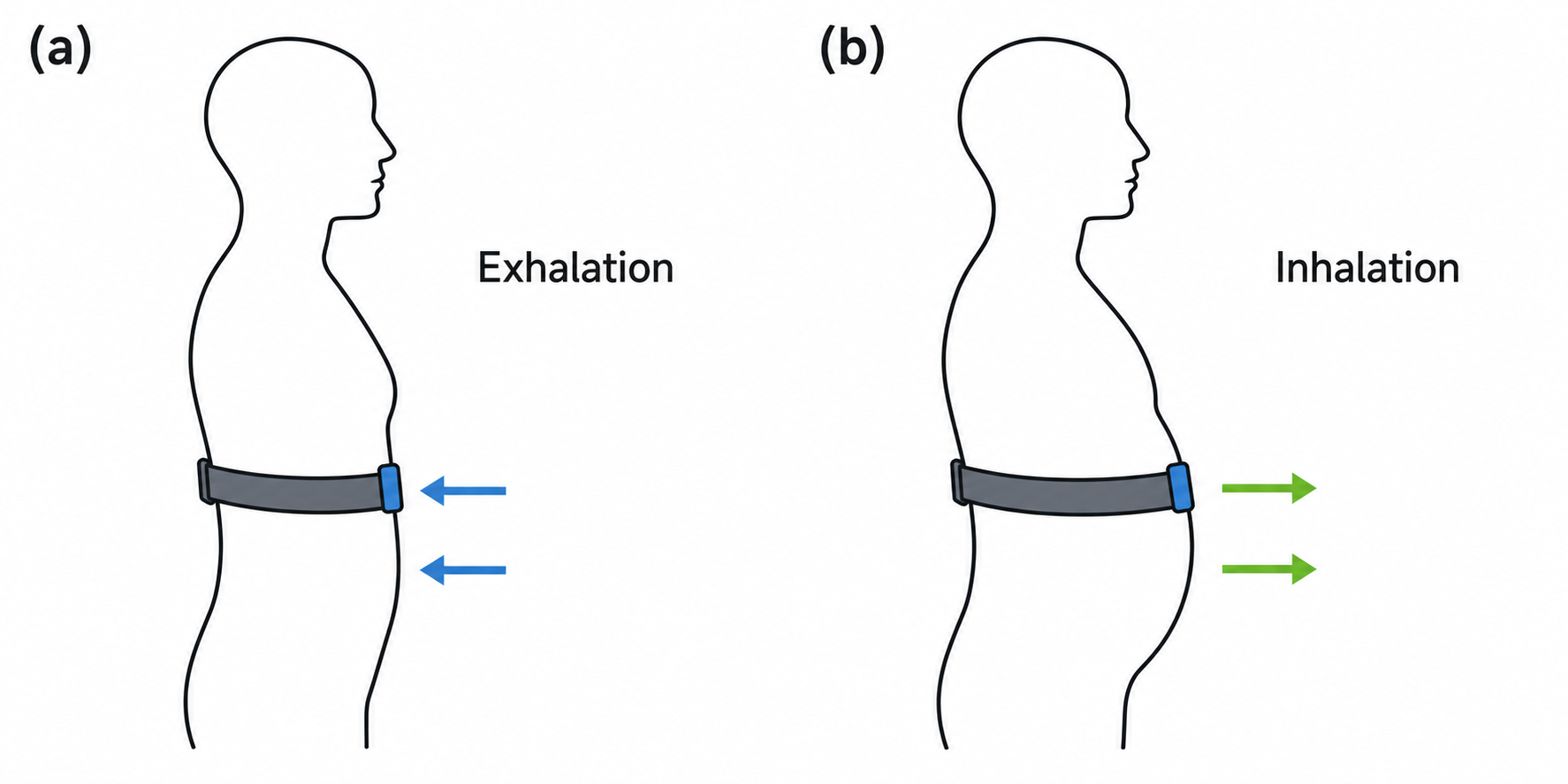}
    \caption{Respiration phases used to reason about pressure transfer: exhalation and inhalation change the abdominal geometry and therefore the pressure applied to the FSR.}
    \label{fig:respiration_mechanism}
\end{figure}

During inhalation, diaphragm motion expands the chest and abdominal regions. The mechanical design must therefore convert this expansion into a concentrated and repeatable pressure on the FSR. Figure~\ref{fig:respiration_mechanism} illustrates the respiratory phases considered in the design. The prototype holder was designed so that both inhalation and exhalation produce a measurable pressure difference at the sensing point, maximizing signal amplitude while preserving comfort.

\subsubsection{Signal Observation and Initial Evaluation}

The Arduino prototype was programmed to sample the analog signal and transmit it for visualization and MATLAB analysis. The raw signals confirmed that the system could capture distinct breathing patterns. With the $5~\mathrm{V}$ Arduino divider supply, the observed voltage swing was large in the feasibility recordings: normal breathing produced at least a $500~\mathrm{mV}$ swing, while deep breathing reached approximately $900~\mathrm{mV}$ (Figure~\ref{fig:raw_modes}). Fast breathing was captured without visible delay, and breath holding produced an approximately constant signal with only minor noise.

Because the FSR is a passive resistive element, its current draw is determined by the divider supply and the instantaneous sensor resistance rather than by an active sensing circuit. Under low-pressure or standby conditions, the high FSR resistance substantially reduces divider current, lowering the sensing element's contribution to standby power and thereby helping extend battery life. The large observed signal swing and simple readout supported the move from the Arduino feasibility prototype to a compact PCB-based wearable system.

\subsection{PCB Design and Embedded System Architecture}
\label{sec:pcb_architecture}

In this section, we describe the design of the integrated printed circuit board (PCB) and the selection of key electronic components used in the proposed wearable respiratory sensing system. The design focuses on achieving low power consumption, compact size, and reliable wireless communication for continuous monitoring.

\subsubsection{Component Selection}

We first introduce the main components used in the system and discuss the rationale behind their selection, along with their role in the overall architecture.

\paragraph{Microcontroller Unit (MCU).}
The central processing unit of the system is the \texttt{nRF52832}, a low-power microcontroller with built-in Bluetooth Low Energy (BLE) communication at 2.4 GHz \citep{nrf52832}. This MCU was selected because it balances computational capability, ADC support, wireless communication, and energy efficiency, which are all critical for continuous wearable operation.

The nRF52832 includes a 12-bit analog-to-digital converter (ADC) with oversampling support, allowing the effective number of bits to reach up to 14 bits. This enables high-resolution digitization of the analog respiratory signal. The MCU supports I2C and SPI for communication with peripheral devices. Its $1.7$--$3.6~\mathrm{V}$ supply range and low-power operating modes make it suitable for battery-powered wearable applications.

\paragraph{Power Management.}
Power efficiency is a critical requirement for continuous respiratory monitoring. Instead of relying on linear regulation, which wastes power as heat, the design uses a switching buck converter to minimize loss. Specifically, the \texttt{LM3670-1.8V} converter provides a stable $1.8~\mathrm{V}$ supply and can achieve efficiency above 90\% under suitable operating conditions \citep{lm3670}. This choice supports longer battery life compared with a low-dropout-only design.

\paragraph{Wireless Communication and Antenna.}
Wireless data transmission is implemented through BLE on the MCU. For RF transmission, the design uses the \texttt{Johanson 2450AT18B100E} chip antenna \citep{johanson2450}. A matching network is placed between the MCU RF output and the antenna to improve impedance matching and transmission efficiency. This RF network is implemented directly on the PCB.

\paragraph{Battery Charging Circuit.}
To support reusability and portability, the system is powered by a rechargeable battery. A dedicated charging IC, \texttt{BQ2409x}, manages safe and efficient charging \citep{bq2409x}. This makes the device usable with a compact battery while avoiding the need for disposable cells.

\paragraph{Micro-USB Interface.}
A micro-USB interface provides external power and battery charging. This allows the device to be charged from common sources such as power banks, laptop USB ports, or USB adapters, improving usability in real-world settings.

\paragraph{Vibration Motor for Feedback.}
A vibration motor is integrated to provide haptic feedback. This feedback can notify the user of low battery or improper placement, for example when excessive pressure is applied to the sensor or when the sensor loses contact. A dedicated motor driver is used to control the motor efficiently.

\subsubsection{Circuit Design and PCB Implementation}

\begin{figure}[t]
    \centering
    \includegraphics[width=0.92\linewidth]{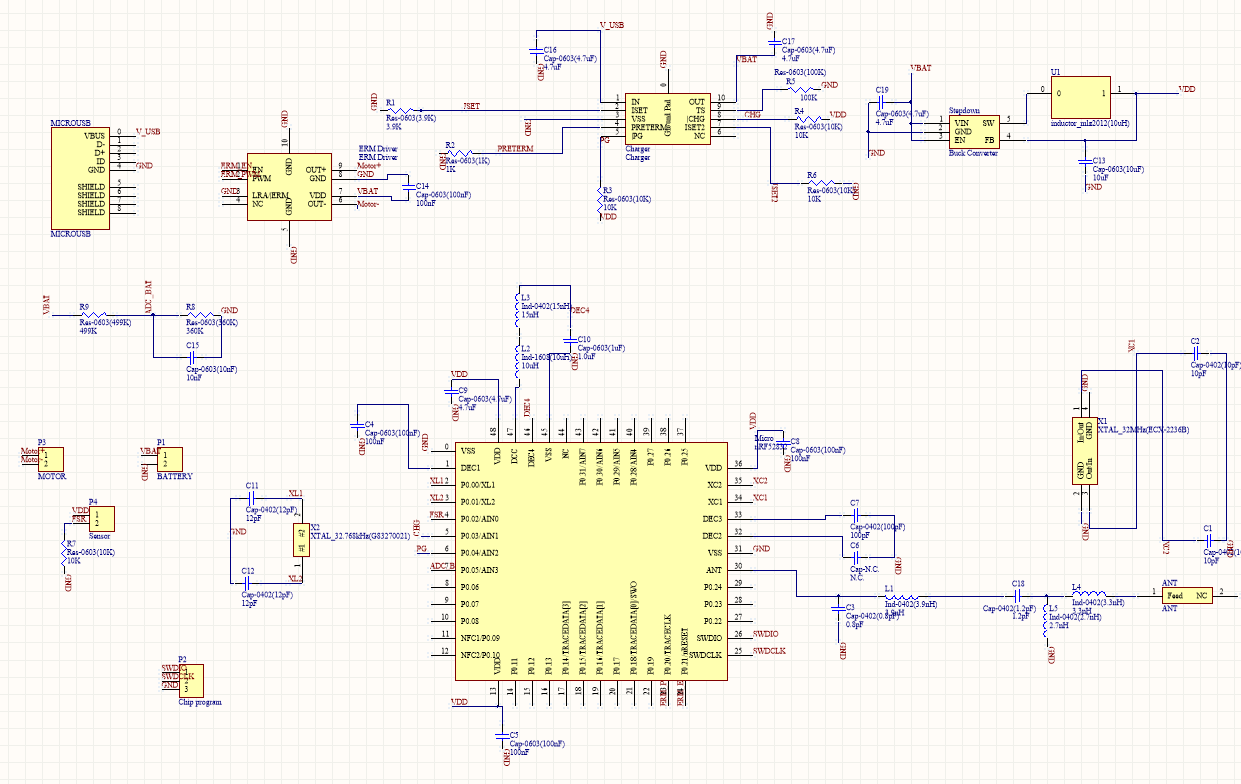}
    \caption{Schematic of the designed electronic system, including the MCU, power-management circuitry, BLE/RF path, charging circuit, and peripheral components.}
    \label{fig:pcb_schematic}
\end{figure}

All components described above were integrated into a unified circuit design using Altium Designer. The schematic is shown in Figure~\ref{fig:pcb_schematic}. During PCB design, particular attention was given to minimizing board size while preserving signal integrity and RF performance. Components were arranged compactly to reduce the wearable footprint and manufacturing cost.

\begin{figure}[t]
    \centering
    \includegraphics[width=0.62\linewidth]{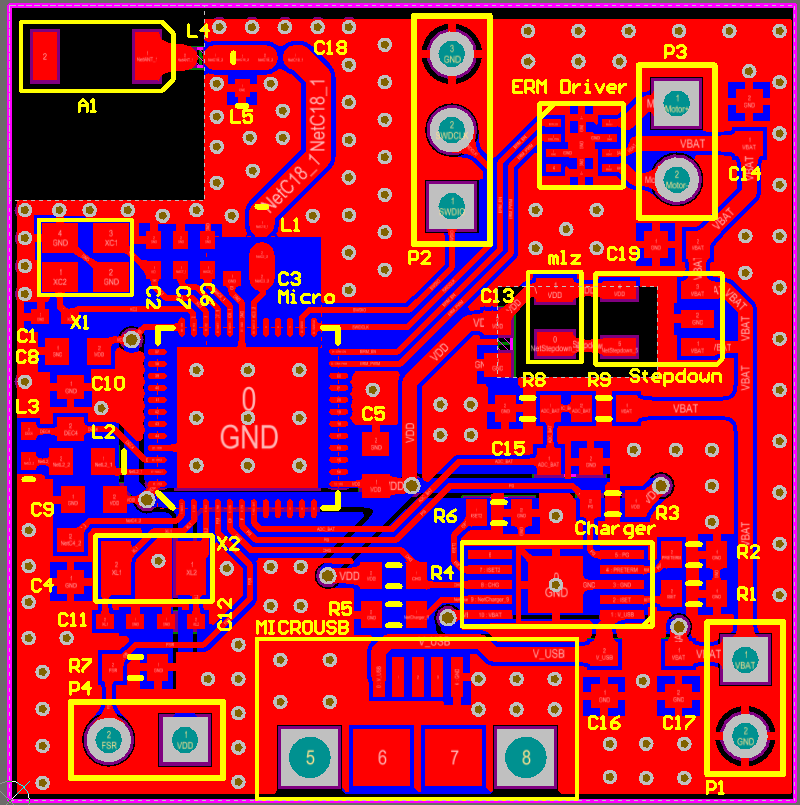}
    \caption{Layout of the designed PCB. Compact placement and polygon planes reduce footprint while supporting stable mixed-signal and RF operation.}
    \label{fig:pcb_layout}
\end{figure}

Two polygon planes were used in the PCB layout. These planes improve thermal dissipation, reduce power-supply fluctuations, and enhance electrical stability. The resulting PCB layout is shown in Figure~\ref{fig:pcb_layout}. The fabricated board measured approximately $2.5~\mathrm{cm}\times2.5~\mathrm{cm}$, making it suitable for integration into a belt-mounted enclosure.

\subsubsection{Fabrication and Assembly}

\begin{figure}[t]
    \centering

    \begin{subfigure}[t]{0.47\linewidth}
        \centering
        \includegraphics[
            width=\linewidth,
            height=5cm,
            keepaspectratio
        ]{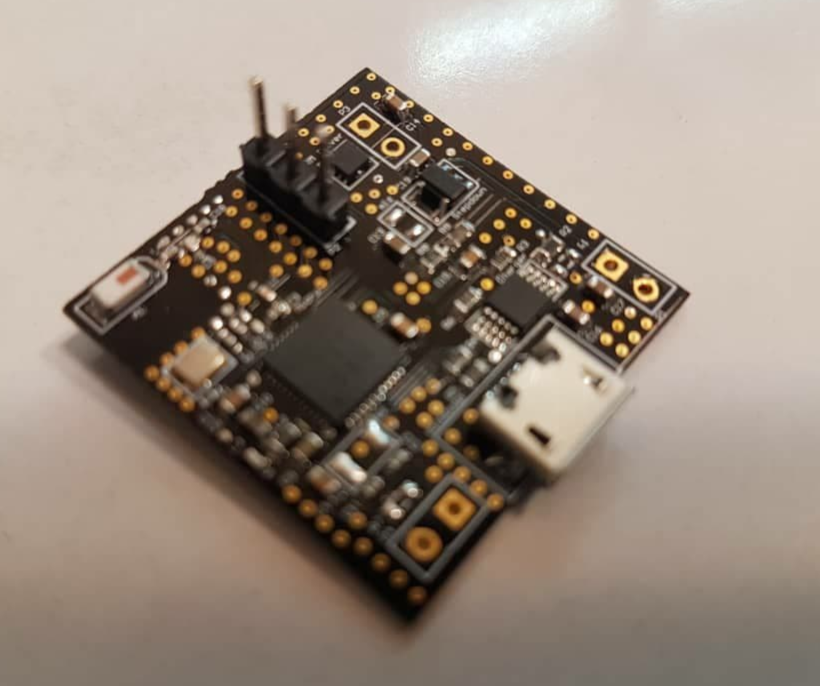}
        \caption{Assembled PCB}
    \end{subfigure}
    \begin{subfigure}[t]{0.47\linewidth}
        \centering
        \includegraphics[
            width=\linewidth,
            height=5cm,
            keepaspectratio
        ]{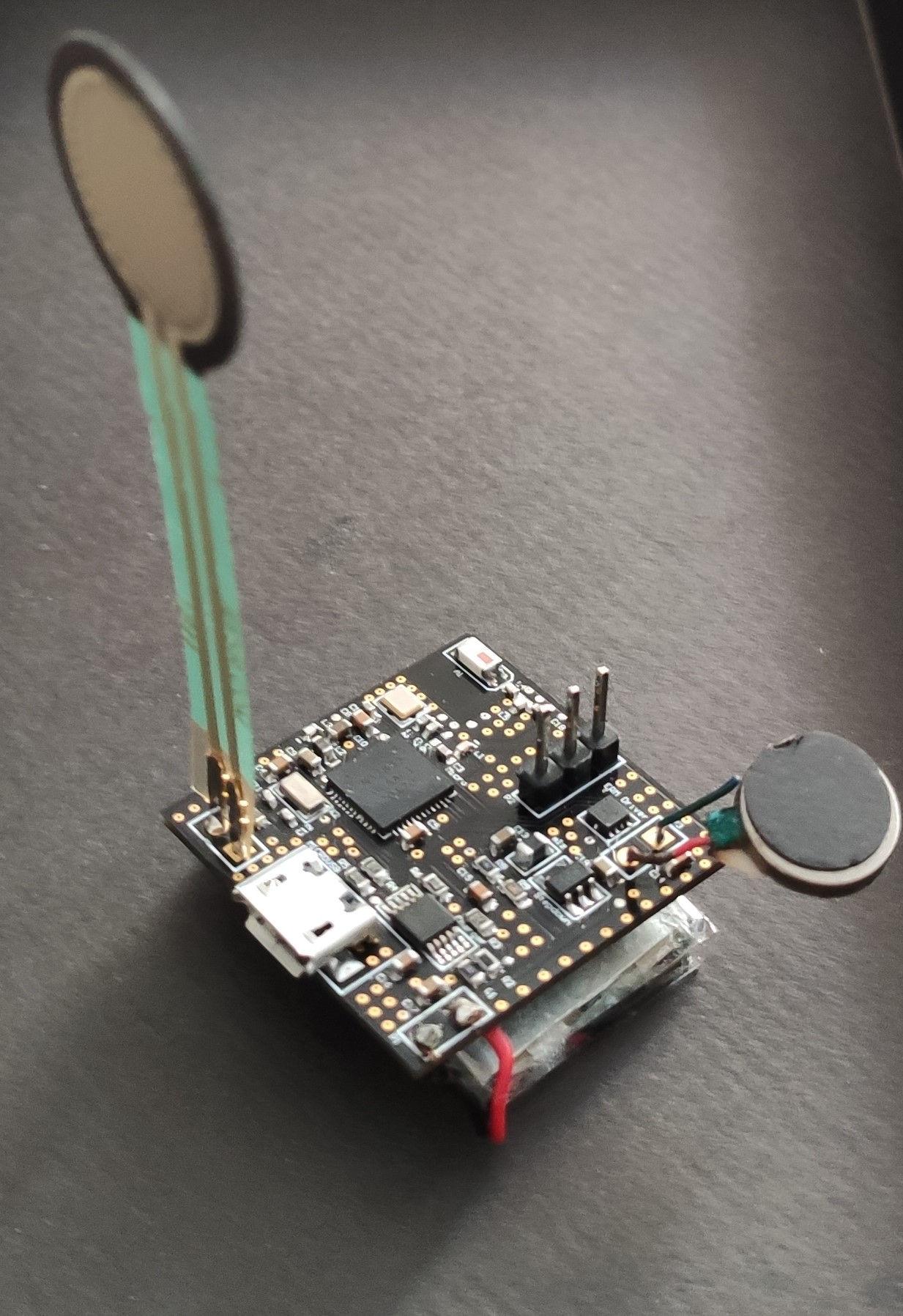}
        \caption{Device with sensor and motor}
    \end{subfigure}

    \caption{Assembled respiratory-sensing electronics. The final board is compact enough for wearable use while integrating acquisition, BLE communication, charging, power management, and feedback circuitry.}
    \label{fig:pcb_assembled}
\end{figure}

After completing the PCB design, the board was fabricated and assembled. The assembled prototype is shown in Figure~\ref{fig:pcb_assembled}. Following assembly, the microcontroller was programmed to acquire sensor data and transmit it wirelessly through BLE to a computer for processing and visualization.

\subsection{Sensor Holder Design and Evaluation}
\label{sec:holder_design}

After implementing the PCB-based system with the FSR sensor, we designed a complete wearable body structure to extract a signal representative of respiration. The holder had to protect the PCB, keep the FSR mechanically stable, and apply pressure in a controlled manner without making the belt uncomfortable.

\subsubsection{Sensor Holder Design}

\begin{figure}[t]
    \centering
    \begin{subfigure}{0.48\linewidth}
        \centering
        \includegraphics[width=\linewidth]{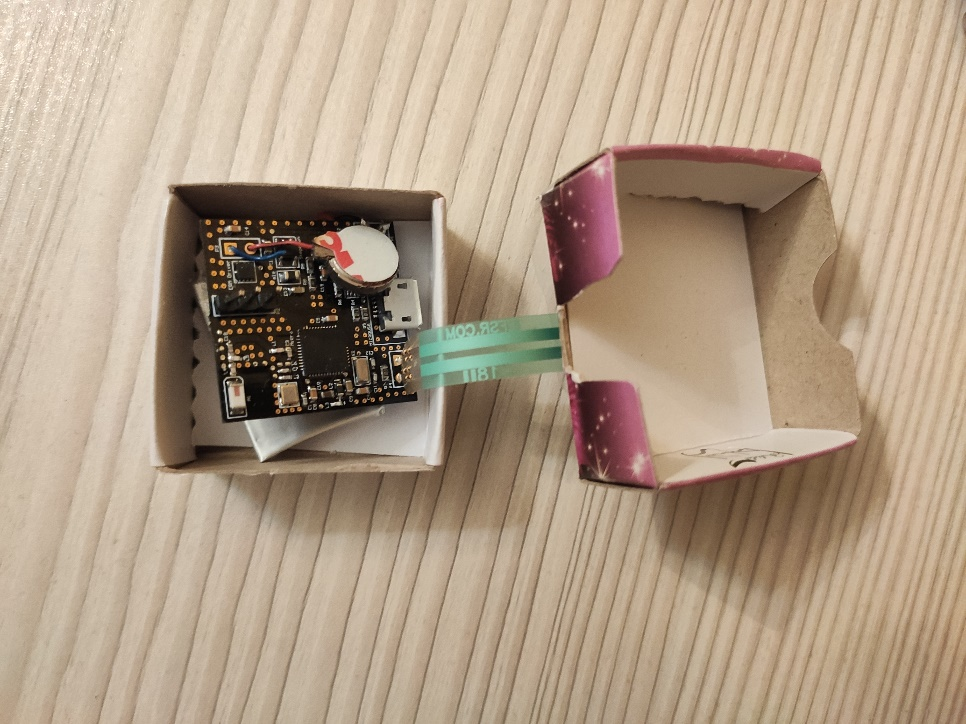}
        \caption{Enclosure}
    \end{subfigure}
    \hfill
    \begin{subfigure}{0.48\linewidth}
        \centering
        \includegraphics[width=\linewidth]{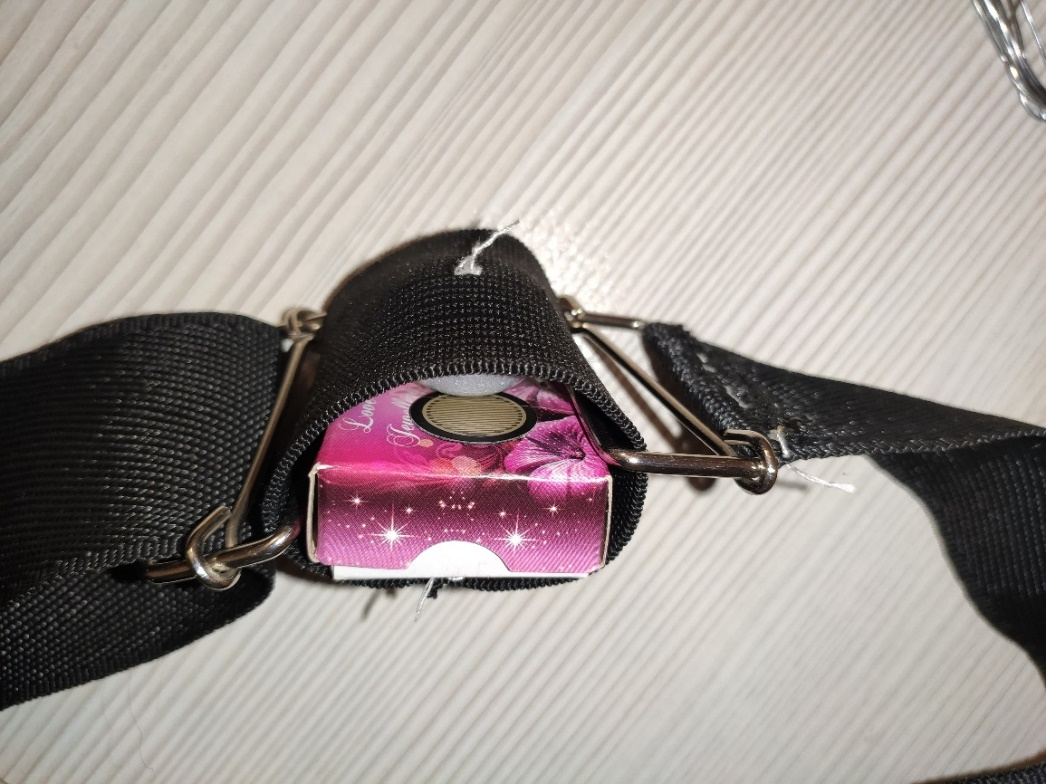}
        \caption{Flexible ring}
    \end{subfigure}
    \hfill
    \par\medskip
    \begin{subfigure}{0.48\linewidth}
        \centering
        \includegraphics[width=\linewidth]{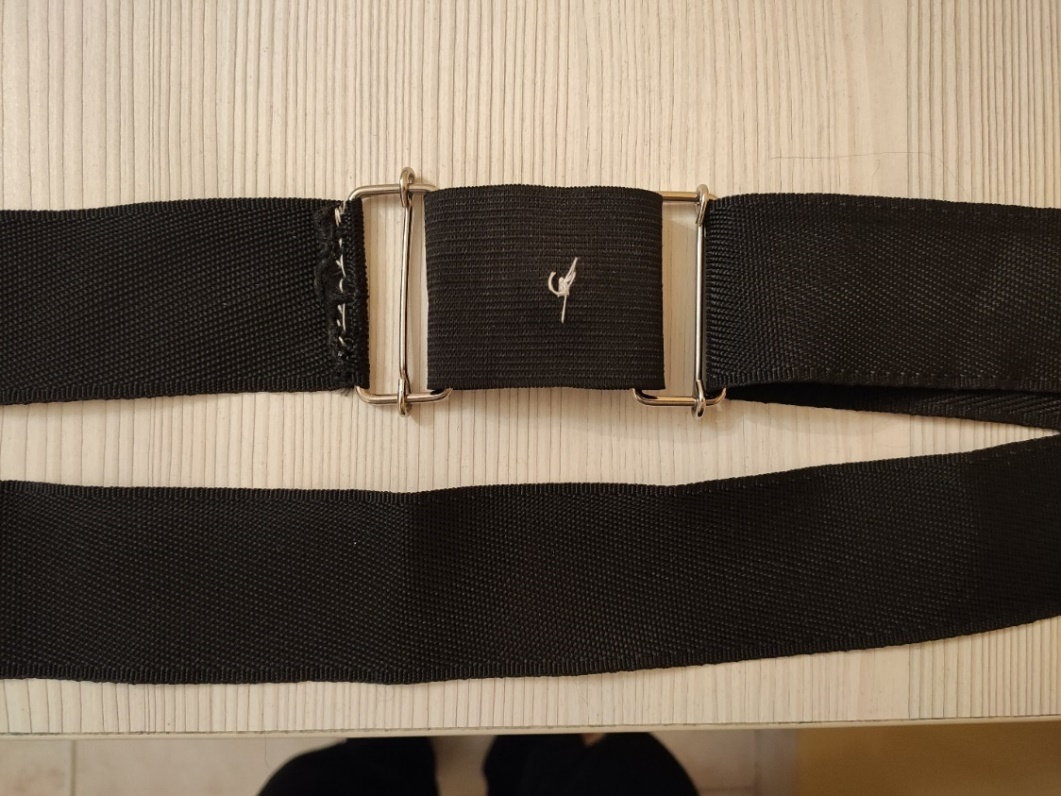}
        \caption{Wearable belt}
    \end{subfigure}
        \hfill
    \begin{subfigure}{0.48\linewidth}
        \centering
        \includegraphics[height=0.672\linewidth,angle=-90,origin=c,keepaspectratio]{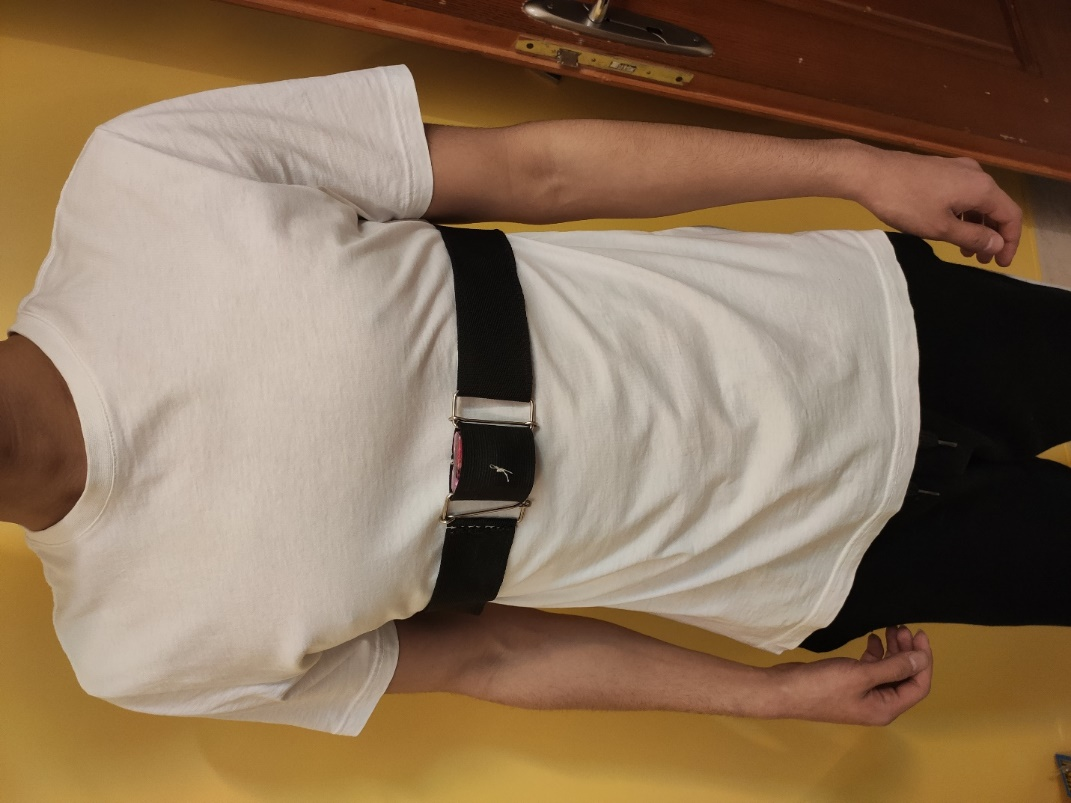}
        \caption{Body placement}
                \label{fig:body_position}

    \end{subfigure}
    \caption{Final wearable holder and placement. The enclosure protects the electronics, the central ring concentrates pressure on the FSR, the side bands stabilize the system, and panel (d) shows the upper-abdominal sensor position.}
    \label{fig:holder_design}
\end{figure}

The main idea was to place the PCB and the FSR inside a protective enclosure (Figure~\ref{fig:holder_design}a) while exposing the FSR at the pressure-transfer point. A small enclosure with dimensions $3.5\times3.5\times1.8~\mathrm{cm}$ was used. The PCB was placed inside this protective box, and the FSR was fixed through the enclosure wall so it remained stable on the outer surface.

A highly flexible strap was wrapped around the enclosure (Figure~\ref{fig:holder_design}b). A button was sewn inside this strap and placed directly on top of the FSR. Tightening the strap caused the button to apply pressure to the FSR, changing its resistance. The flexible ring was then fixed between two metal clips and attached to two elastic bands with lower stiffness than the central strap (Figure~\ref{fig:holder_design}c). These bands secured the system around the waist and allowed natural breathing motion.

\subsubsection{Sensor Placement on the Body}

The movement of the diaphragm and abdomen provides a measurable mechanical surrogate for respiratory activity, although the mapping depends on subject anatomy, posture, and placement \citep{chu2019respiration,chen2021individuality}. To obtain a signal correlated with breathing activity, the sensor was placed on the upper abdominal region. Placing the sensor higher, closer to the chest, reduced the applied pressure and therefore lowered the voltage swing. However, higher placement also reduced sensitivity to some walking-related motion patterns. The final position therefore reflects a trade-off between signal amplitude and motion sensitivity, as illustrated in Figure~\ref{fig:body_position}.

\subsubsection{Signal Acquisition in Different Conditions}
\label{sec:condition_acquisition}

In the final validation stage, respiratory signals were recorded at different body positions. The collected data from the PCB were transmitted to MATLAB using BLE.

To reduce motion artifacts during respiration measurement, digital filtering techniques were applied. A finite impulse response (FIR) low-pass filter with a Hamming window was used, with a 3 dB cutoff frequency of $2~\mathrm{Hz}$. Adult resting respiratory rates are commonly about 12--20 breaths per minute, well below the $2~\mathrm{Hz}$ cutoff; the filter therefore retains the respiratory waveform and faster voluntary breathing while attenuating higher-frequency noise \citep{nicolo2020importance}.

The raw data acquired from the PCB also undergo internal digital averaging in the microcontroller ADC stage. Oversampling was used with a factor of 8, meaning that every 8 raw samples produce one averaged output value. The sampling rate was set to $428~\mathrm{Hz}$, which is much higher than the respiratory frequency range and was mainly used for testing and validation. The BLE packet interval was set to $17.5~\mathrm{ms}$; each packet transmitted 8 ADC samples to MATLAB. With a 14-bit ADC output stored as 2 bytes per sample, each packet carried 16 bytes, corresponding to approximately $914$ payload bytes per second, or $7.31~\mathrm{kbps}$.

Data acquisition was performed in four body positions: sitting, lying down, standing, and walking. In each condition, four respiratory states were analyzed: normal breathing, deep breathing, apnea or breath holding, and fast breathing. In total, measurements were performed across 16 experimental condition-mode combinations (Figure~\ref{fig:posture_trials}).

\paragraph{Sitting condition.}
Respiratory signals were recorded for 2 minutes while the subject was sitting and leaning back on a chair. The filtered signal and raw signal were compared, and four breathing patterns were extracted: normal, deep, breath holding, and fast breathing. Respiratory rate was computed using the time difference between consecutive peaks and valleys, and the resulting waveforms showed clear differences between breathing intensities.

\paragraph{Lying condition.}
Experiments were also conducted in a lying position. A 2-minute recording was performed while the subject was lying on a bed. The signal preserved clear peak-valley structure across the different breathing modes, indicating that the holder can maintain pressure coupling even when the body posture changes substantially.

\paragraph{Standing condition.}
For the standing condition, respiratory data were recorded for approximately 135 seconds. Peaks and valleys corresponding to different breathing patterns remained identifiable after filtering, showing that the system was not limited to seated or supported postures.

\paragraph{Walking condition.}
Respiratory signals were recorded during walking for approximately 130 seconds. Motion artifacts were more pronounced in this condition. However, after filtering, respiratory patterns remained distinguishable in the tested setup. This result is important because it shows partial robustness to natural movement, while also motivating the motion-artifact analysis in the appendix.

Overall, the PCB-based system could measure respiratory rate and intensity across the tested body positions. The observed voltage swing was approximately $500~\mathrm{mV}$ during normal breathing and increased up to roughly $1~\mathrm{V}$ during deep breathing. This large amplitude helps separate respiratory structure from moderate motion artifacts and validates the use of the calibrated respiratory signal in the stress experiment.

\subsection{Stress Experiment Setup}
\label{sec:stress_setup}

To evaluate the potential value of the proposed respiratory sensor for a downstream physiological task, we recorded respiratory signals during alternating rest and stress-induction phases. The resulting dataset was used for a preliminary binary classification of rest versus stress-induction phases.

\subsubsection{Experimental Protocol}
\label{sec:stress_protocol}

The experiment consisted of five labeled stages and lasted approximately 16 minutes and 15 seconds in total. Before the main protocol began, a short setup phase was used to place the sensor on the participant's body, verify the attachment location, and check the stability of the BLE link and data transmission.

During the main protocol, the participant first completed a baseline rest period, followed by two mental-arithmetic stress tasks and one Stroop Color--Word Task (SCWT), with a second rest period at the end. These tasks are commonly used to elicit acute cognitive load and measurable psychophysiological responses \citep{stroop1935studies,tulen1989characterization,cipresso2019computational,scarpina2017stroop}. After each stage, the participant completed a short self-report questionnaire indicating the perceived stress level for that stage. Table~\ref{tab:stress_timeline} summarizes the timing.

\begin{table}[t]
    \centering
    \caption{Timing and sequence of the stress experiment. A short self-report questionnaire was completed after each phase.}
    \label{tab:stress_timeline}
    \begin{tabular}{p{0.22\linewidth}p{0.16\linewidth}p{0.52\linewidth}}
        \toprule
        Phase & Duration & Description \\
        \midrule
        Setup & 1 min & Sensor placement, attachment check, and BLE verification \\
        Rest 1 & 4 min & Baseline breathing under normal conditions \\
        Mental arithmetic 1 & 2 min & Backward counting task with changing starts and steps \\
        SCWT & 3 min & Color naming followed by word reading \\
        Mental arithmetic 2 & 2 min & Second backward-counting task \\
        Rest 2 & 3 min & Recovery and return to baseline breathing \\
        \bottomrule
    \end{tabular}
\end{table}

\subsubsection{Participant Instructions and Self-Report}
\label{sec:stress_self_report}

Before the protocol, each participant was given a short explanation of the task structure, including the mental arithmetic stage, the SCWT stage, and the questionnaire procedure. After each phase, the participant self-reported perceived stress on a five-point scale ranging from ``not at all'' to ``very much''. The full questionnaire is provided in Appendix~\ref{app:questionnaire}.

\subsubsection{Stress-Inducing Tasks}
\label{sec:stress_tasks}

The stress stage consisted of two mental-arithmetic blocks and one SCWT block. In the first mental-arithmetic block, the participant was asked to count backward as quickly as possible from 2047 in steps of 17. Whenever an error occurred, a new starting number and step size were chosen to prevent adaptation to the task.

Next, the participant completed an SCWT block. In the first part, the participant named the colors of printed words. In the second part, the participant read the written English words. Finally, the participant completed a second mental-arithmetic block for 2 minutes, again followed by a short self-report.

\subsubsection{Participants and Recorded Dataset}
\label{sec:stress_participants}

The experiment was conducted on 12 participants, including 4 women and 8 men, with ages ranging from 15 to 50 years. The mean age was 25.3 years. For each participant, we recorded the respiratory signal across the five experimental phases, resulting in 60 labeled segments. The classification labels were assigned from the protocol phase: the two rest segments were labeled rest (0), and the two mental-arithmetic segments and the SCWT segment were labeled stress induction (1).

\subsubsection{Feature Extraction}
\label{sec:stress_features}

After recording, we extracted 28 time-domain features from each respiratory segment. These features describe peak-to-peak structure, valley-to-valley structure, slope statistics, inspiratory and expiratory timing, and amplitude-related properties. The complete feature list is provided in Appendix~\ref{app:features}.

For model training, six features were selected using Fisher's criterion \citep{fisher1936use} because they provided the strongest separation between stress and rest classes: variance of valley intervals, maximum peak-to-valley interval, variance of inspiratory time, mean expiratory volume, variance of expiratory volume, and variance of expiratory slope. Among these, variance of expiratory slope was the most discriminative feature, suggesting that nonlinear variations in the expiratory phase were strongly associated with the stress condition in this dataset.

\subsubsection{Classification Setup}
\label{sec:stress_classification_setup}

We used the extracted features to train standard machine-learning classifiers for binary stress-phase recognition. The models were evaluated using six-fold cross-validation over the 60 labeled segments. In each fold, 50 samples were used for training and 10 samples were used for testing. We evaluated linear discriminant analysis (LDA), $k$-nearest neighbors (KNN), support vector machine (SVM), decision tree, and a one-hidden-layer neural network with 20 hidden units, representing standard linear, local, margin-based, tree-based, and neural baselines.

%% file: results.tex
\section{Experiments and Results}

\subsection{Posture and Motion Evaluation}
\label{sec:posture_motion_results}

Before evaluating posture and motion, we verified that the prototype captures the expected temporal differences among controlled breathing patterns. The corresponding waveforms and analysis are provided in Appendix~\ref{app:prototype_waveforms} (Figure~\ref{fig:raw_modes}).

\begin{figure}[t]
    \centering
    \includegraphics[width=0.92\linewidth]{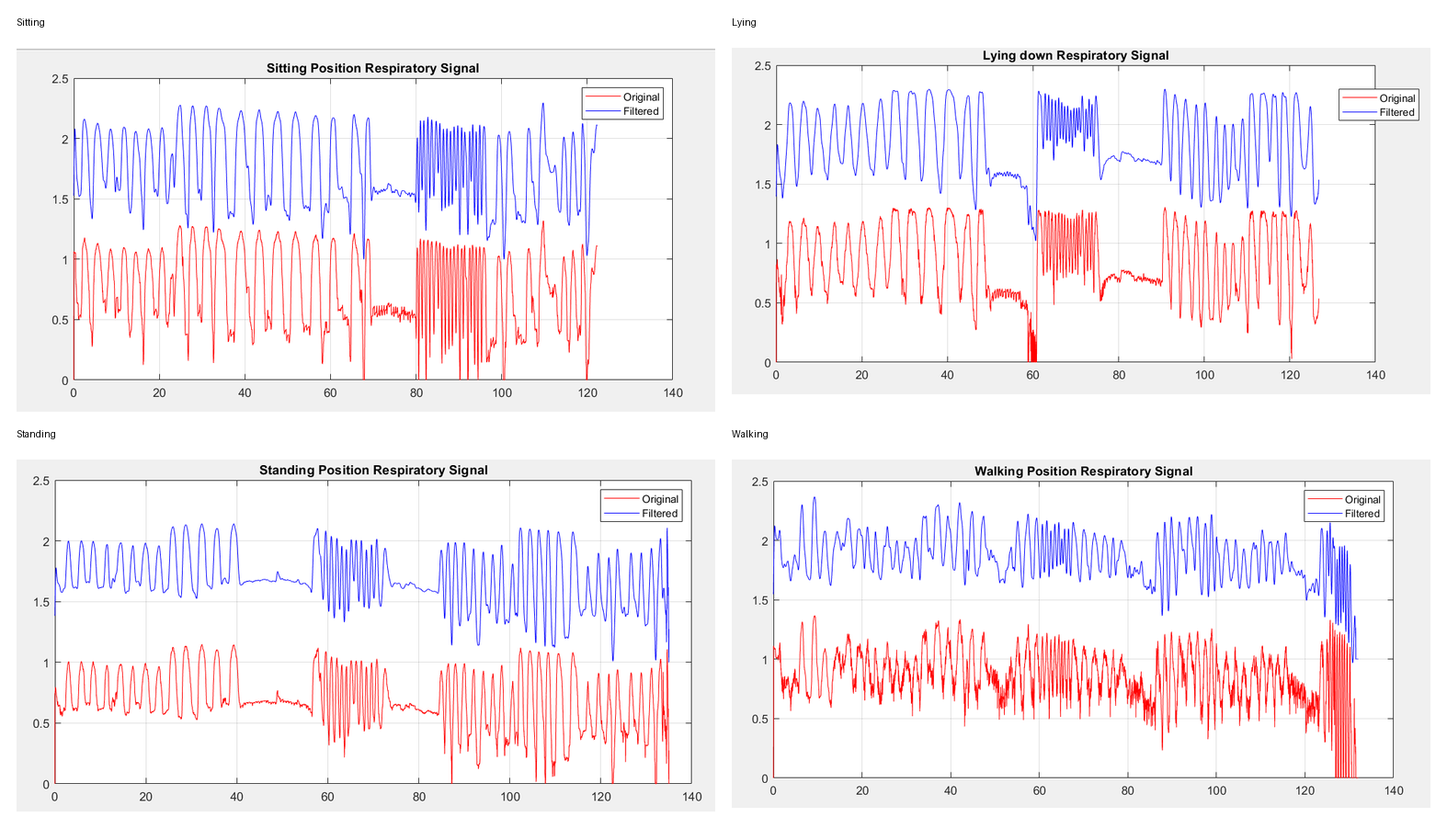}
    \caption{Respiratory recordings across posture and light-motion conditions. Each panel shows filtered and raw traces from the validation experiments.}
    \label{fig:posture_trials}
\end{figure}

We next investigate whether the PCB-based wearable retains respiratory structure as posture and movement change. Figure~\ref{fig:posture_trials} summarizes the raw and filtered signals recorded while sitting, lying, standing, and walking. Clear respiratory cycles remain after filtering in the three stationary postures. Walking introduces stronger low-frequency artifacts, but respiratory structure remains visible for the tested subject, consistent with the known sensitivity of wearable respiratory systems to posture, attachment, and locomotion \citep{zhang2014adaptive,chu2019respiration,chen2021individuality}. Across these trials, normal breathing typically produces a voltage swing of approximately $500~\mathrm{mV}$, while deep breathing can approach $1~\mathrm{V}$. Motion-related fluctuations are often below $200~\mathrm{mV}$ in the controlled trials, although Appendix~\ref{app:motion} presents conditions in which motion and placement dominate the signal.

These results demonstrate that the wearable is not restricted to a single static posture and retains recurring respiratory-scale variations under light movement. They also delineate the operating range of the current design, beyond which placement changes and motion artifacts become the dominant source of variation.

\subsection{Stress Decoding Results}
\label{sec:stress_results}

\subsubsection{Respiratory Signal Patterns Across Task Phases}
\label{sec:stress_signal_results}

\begin{figure}[t]
    \centering
    \includegraphics[width=0.92\linewidth]{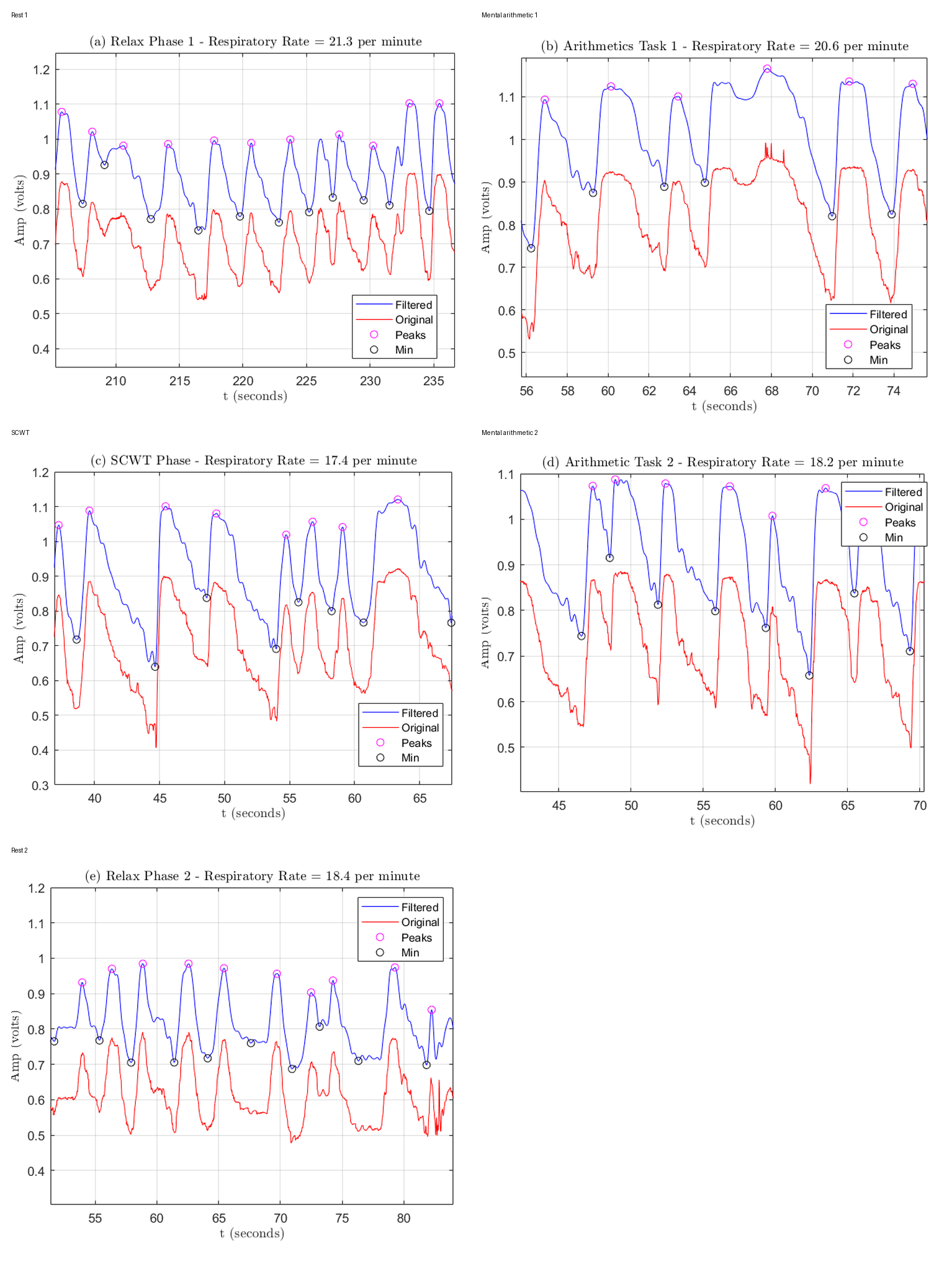}
    \caption{Respiratory signal from one participant across the five stages of the stress experiment: first rest, first mental-arithmetic block, SCWT, second mental-arithmetic block, and second rest.}
    \label{fig:stress_signal_phases}
\end{figure}

We next examine how the measured respiratory waveform changes across the five phases of the stress experiment. Figure~\ref{fig:stress_signal_phases} shows a representative recording from one participant. Compared with the resting periods, the waveform changes noticeably during the mental-arithmetic and SCWT phases. Participants also speak during these tasks, directly altering the respiratory pattern relative to the resting baseline.

In particular, speaking can lengthen inhalation and increase the time spent near local peaks or valleys. The resulting waveform is less regular and exhibits a smoother downward slope. These observations are consistent with prior evidence that cognitive demand, emotion, and speech production can alter respiratory timing and variability \citep{boiten1998effects,grassmann2016respiratory,homma2008breathing,conrad1979speech}. Thus, the recorded differences contain contributions from both the intended stress manipulation and the behavior required by the tasks.

The distinct phase-wise structure shows that the proposed wearable captures task-associated respiratory changes that are sufficiently pronounced to support downstream feature-based decoding, while also motivating a careful interpretation of what those features represent.

\subsubsection{Cold Pressor Test}
\label{sec:cold_pressor_results}

\begin{figure}[t]
    \centering
    \includegraphics[width=0.84\linewidth]{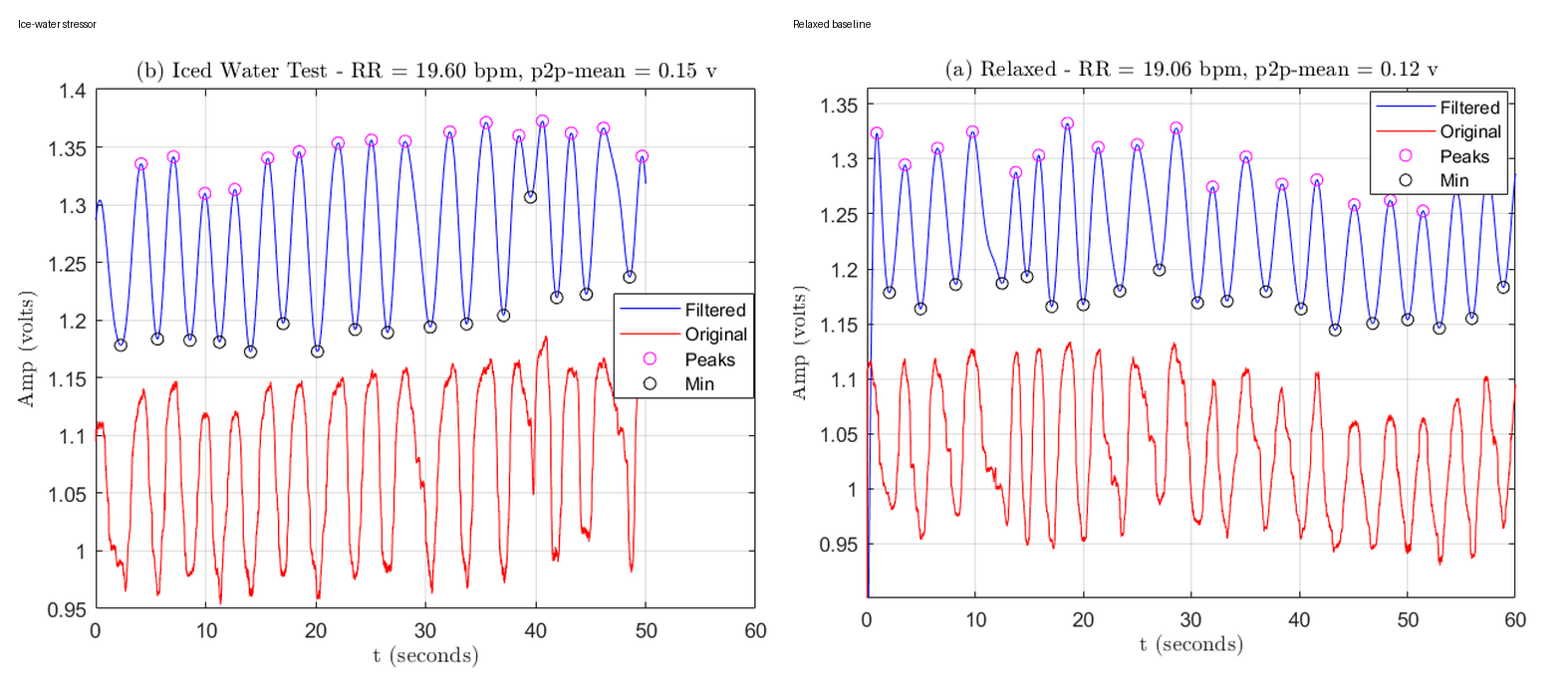}
    \caption{Cold pressor test: baseline rest and hand immersion in ice water. This task was explored as an alternative stress induction method.}
    \label{fig:cold_pressor_test}
\end{figure}

Before selecting the main protocol, we also explored the cold pressor test, a standard autonomic stressor in which the participant immerses one hand in ice water \citep{lovallo1975cold}. In our preliminary test, respiratory signal amplitude increased by approximately 20\%. However, this change did not yield a sufficiently clear or stable signature for the intended data-collection setting. Because the test is also painful and difficult to sustain for a prolonged period, we did not retain it as the primary stress-induction protocol and instead used the mental-arithmetic and SCWT tasks. Figure~\ref{fig:cold_pressor_test} illustrates the exploratory cold pressor setup.

This exploratory result shows that a standard physiological stressor does not necessarily produce a sufficiently stable respiratory signature under the practical constraints of the experiment, and it directly motivates our final protocol choice.

\subsubsection{Binary Classification Performance}
\label{sec:stress_classification_results}

The same six selected respiratory features were extracted from every phase segment. The binary target, however, was assigned from the experimental protocol rather than from a feature value or questionnaire threshold. The corresponding phase-to-class mapping is provided in Appendix Table~\ref{tab:feature_example}. This labeling makes the reported task explicitly binary while retaining the same interpretable respiratory feature representation for every segment.

We then assess whether these respiratory features distinguish the binary rest and stress-induction classes using five standard classifiers. Table~\ref{tab:stress_classification} reports the resulting sensitivity, specificity, and accuracy. We find that the one-hidden-layer neural network achieves the highest test accuracy of 88.0\%, while the average test accuracy across all evaluated models is 82.4\%.

\begin{table}[t]
    \centering
    \caption{Binary stress-phase classification results for the evaluated models. The reported values follow the metrics used in the original analysis.}
    \label{tab:stress_classification}
    \resizebox{\linewidth}{!}{%
    \begin{tabular}{lcccccc}
        \toprule
        Model & Test spec. & Train spec. & Test sens. & Train sens. & Test acc. & Train acc. \\
        \midrule
        LDA & 0.933 & 0.650 & 0.907 & 1.000 & 0.820 & 0.944 \\
        KNN & 0.863 & 0.650 & 0.933 & 0.700 & 0.780 & 0.840 \\
        SVM & 0.867 & 0.800 & 0.953 & 0.870 & 0.840 & 0.920 \\
        Decision tree & 0.767 & 0.850 & 0.907 & 0.960 & 0.800 & 0.928 \\
        One-hidden-layer NN & 0.933 & 0.800 & 0.960 & 0.960 & 0.880 & 0.948 \\
        \bottomrule
    \end{tabular}}
\end{table}

Within this segment-level evaluation, the neural network performs best, and the variance of expiratory slope is the most discriminative selected feature. These results demonstrate that the acquired respiratory waveform contains information that separates the phases of the present protocol. However, the small sample and the confounding effects of speech and task behavior do not establish that the model or selected features will generalize across subjects, a concern also emphasized in larger wearable studies \citep{hovsepian2015cstress,schmidt2018wesad,chen2021individuality}.
Finally, we place the preliminary classification results in the context of prior stress-detection studies in Appendix~\ref{app:baseline}. The comparison summarizes the number of participants and features, classifier type, sensing modalities, and reported accuracy for each study.

Our results show that even a single low-cost respiratory channel contains useful information about the experimental task phases. Because the compared studies differ in cohorts, labels, validation procedures, and sensor sets, their accuracies should not be interpreted as a controlled ranking. Instead, the comparison supports the feasibility of the sensing approach and motivates a larger subject-independent evaluation.

%% file: discussion.tex
\section{Discussion}

We develop a compact FSR-based wearable designed for low-energy, low-complexity respiratory monitoring and integrate the complete path from mechanical pressure transfer and passive deformation sensing to a custom BLE PCB, wireless acquisition, signal processing, and downstream inference. Across controlled breathing modes, the device produces high-amplitude waveforms with the expected changes in periodicity and amplitude, while the PCB-based system preserves recognizable respiratory structure across stationary postures and light movement. The preliminary decoding experiment further demonstrates the feasibility of using features extracted from these waveforms to distinguish relaxation from stress-induction phases in the present protocol.

The acquisition path couples a passive FSR and voltage-divider readout with a mechanically focused holder and compact BLE board, enabling direct ADC sampling without analog amplification and keeping the electronics simple for wearable use \citep{massaroni2019contact,chu2019respiration,hussain2023wearable}. The holder and belt are integral to performance: they transfer small abdominal deformations to the sensor, while attachment pressure and placement set the operating point and can produce baseline shifts or contact artifacts. Across sitting, lying, standing, and light walking, the system retains usable respiratory structure in representative daily conditions; sensor fusion or accelerometer-assisted filtering may further reduce locomotion artifacts, but stable mechanical contact remains essential \citep{zhang2014adaptive,chen2021individuality,comeau2026aienabled}.

The stress-decoding findings remain preliminary because the dataset contains 60 phase-labeled segments from 12 participants, the original segment-level evaluation is limited, and speech and task behavior directly alter respiration during the arithmetic and SCWT phases. Because respiration also varies with cognitive load and emotion more broadly \citep{boiten1998effects,grassmann2016respiratory,homma2008breathing}, the current result demonstrates discrimination between relaxation and stress-induction phases within this protocol, while stronger claims will require larger cohorts, held-out-participant evaluation, repeated sensor placements, and synchronized reference measurements \citep{healey2005detecting,hovsepian2015cstress,schmidt2018wesad,gjoreski2016continuous}. Richer temporal models could better use waveform dynamics \citep{mcclure2020classification,comeau2026aienabled}; however, respiration alone may not provide sufficient stress specificity, and combining it with electrodermal activity, ECG/HRV, and motion sensing offers a more promising path toward robust multimodal stress inference \citep{healey2005detecting,hovsepian2015cstress,schmidt2018wesad}.

\newpage

%% file: appendix.tex
\section{Appendix}

\section{Questionnaire Used in the Stress Experiment}
\label{app:questionnaire}

After each experimental phase, participants completed the short self-report questionnaire shown in Table~\ref{tab:questionnaire}, describing their perceived stress level.

\begin{table}[t]
    \centering
    \caption{Self-report questionnaire used after each phase of the stress experiment.}
    \label{tab:questionnaire}
    \begin{tabular}{p{0.42\linewidth}p{0.48\linewidth}}
        \toprule
        Question / phase & Response scale \\
        \midrule
        End of first rest period & Not at all / A little / Somewhat / Much / Very much \\
        End of first mental-arithmetic task & Not at all / A little / Somewhat / Much / Very much \\
        End of SCWT & Not at all / A little / Somewhat / Much / Very much \\
        End of second mental-arithmetic task & Not at all / A little / Somewhat / Much / Very much \\
        End of second rest period & Not at all / A little / Somewhat / Much / Very much \\
        \bottomrule
    \end{tabular}
\end{table}

\section{Binary Stress-Phase Label Assignment}
\label{app:binary_labels}

Table~\ref{tab:feature_example} reports the phase-to-class mapping used for the binary classification experiment.

\begin{table}[t]
    \centering
    \caption{Binary labels assigned to the five phases of the stress experiment.}
    \label{tab:feature_example}
    \begin{tabular}{p{0.42\linewidth}p{0.42\linewidth}}
        \toprule
        Phase & Binary class label \\
        \midrule
        Rest 1 & Rest (0) \\
        Mental arithmetic 1 & Stress induction (1) \\
        SCWT & Stress induction (1) \\
        Mental arithmetic 2 & Stress induction (1) \\
        Rest 2 & Rest (0) \\
        \bottomrule
    \end{tabular}
\end{table}

\section{Complete Respiratory Feature Set}
\label{app:features}

We extracted 28 respiratory features from each segment. For readability, we group them into four categories.

\subsection{Peak/Valley Timing Features}
\begin{itemize}
    \item Mean peak interval
    \item Variance of peak interval
    \item Mean valley interval
    \item Variance of valley interval
    \item Minimum peak-to-valley interval
    \item Maximum peak-to-valley interval
\end{itemize}

\subsection{Inspiratory and Expiratory Timing Features}
\begin{itemize}
    \item Mean inspiratory time
    \item Variance of inspiratory time
    \item Mean expiratory time
    \item Variance of expiratory time
\end{itemize}

\subsection{Amplitude / Depth Features}
\begin{itemize}
    \item Mean respiratory amplitude
    \item Variance of respiratory amplitude
    \item Mean inspiratory depth
    \item Variance of inspiratory depth
    \item Mean expiratory depth
    \item Variance of expiratory depth
    \item Mean inspiratory volume
    \item Variance of inspiratory volume
    \item Mean expiratory volume
    \item Variance of expiratory volume
\end{itemize}

\subsection{Slope and Derivative Features}
\begin{itemize}
    \item Mean maximum slope
    \item Variance of maximum slope
    \item Mean inspiratory slope
    \item Variance of inspiratory slope
    \item Mean expiratory slope
    \item Variance of expiratory slope
    \item Mean derivative peak spacing
    \item Variance of derivative peak spacing
\end{itemize}

\section{Prototype Respiratory Waveforms}
\label{app:prototype_waveforms}
\label{sec:waveform_quality}

\begin{figure}[t]
    \centering
    \includegraphics[width=0.92\linewidth]{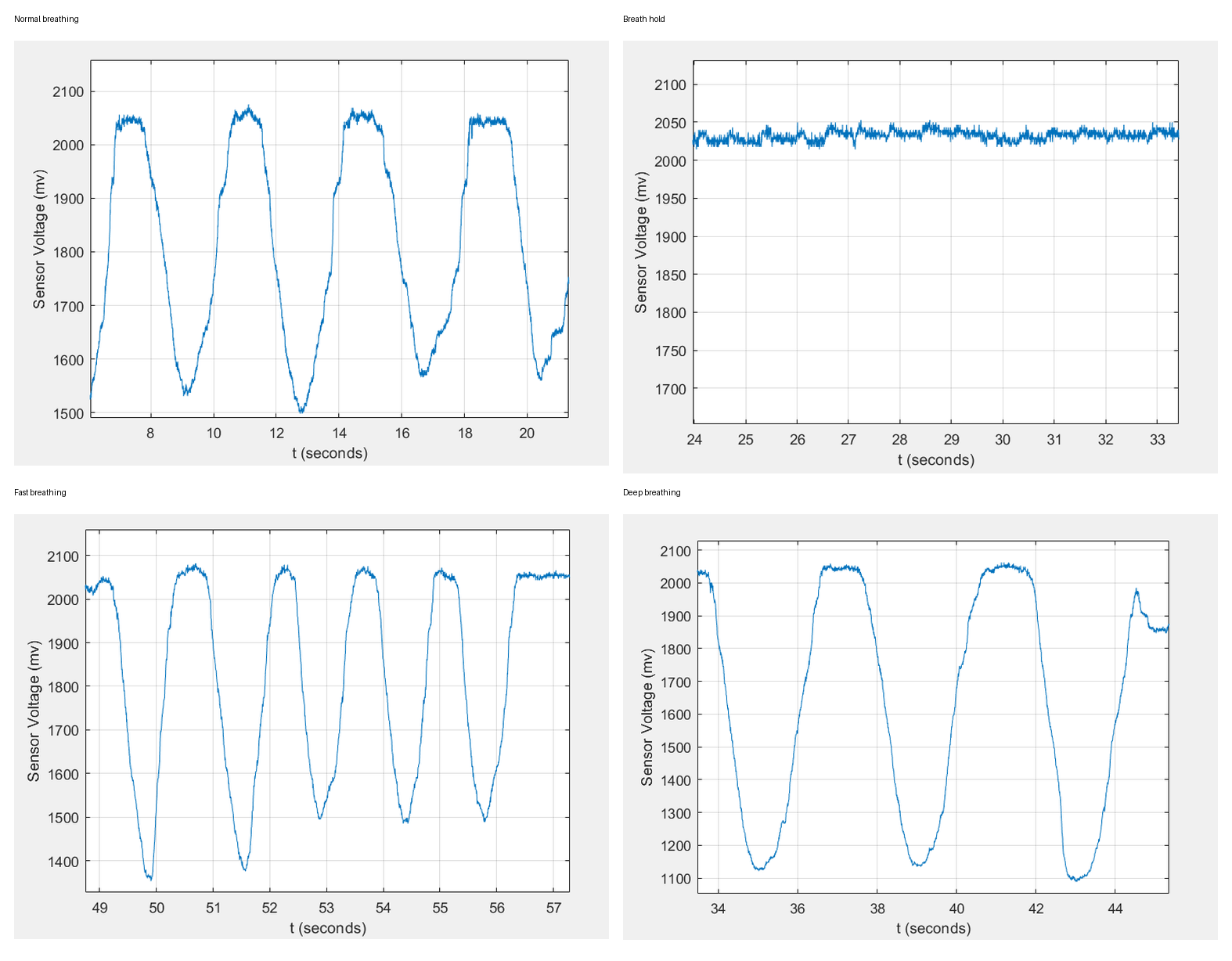}
    \caption{Representative prototype waveforms for normal breathing, breath holding, fast breathing, and deep breathing. The large voltage swing allowed direct ADC sampling without analog amplification.}
    \label{fig:raw_modes}
\end{figure}

We evaluate whether the prototype captures the expected temporal structure of different breathing patterns. Figure~\ref{fig:raw_modes} shows representative signals during normal breathing, breath holding, fast breathing, and deep breathing. We find that normal breathing produces a repeatable periodic waveform, while breath holding largely removes this periodic component. Fast breathing decreases the intervals between successive peaks and valleys, whereas deep breathing produces a larger signal amplitude.

\FloatBarrier
\FloatBarrier

Because the FSR is a passive resistive element, its current draw is determined by the divider supply and the instantaneous sensor resistance rather than by an active sensing circuit. Under low-pressure or standby conditions, the high FSR resistance substantially reduces divider current, lowering the sensing element's contribution to standby power and thereby helping extend battery life. The large observed signal swing and simple readout supported the move from the Arduino feasibility prototype to a compact PCB-based wearable system. Together, these recordings show that the core sensing mechanism preserves interpretable differences among controlled respiratory behaviors and provides a sufficiently large signal for direct digitization without analog amplification.

\section{Motion Artifacts}
\label{app:motion}

\begin{figure}[h]
    \centering
    \includegraphics[width=0.92\linewidth]{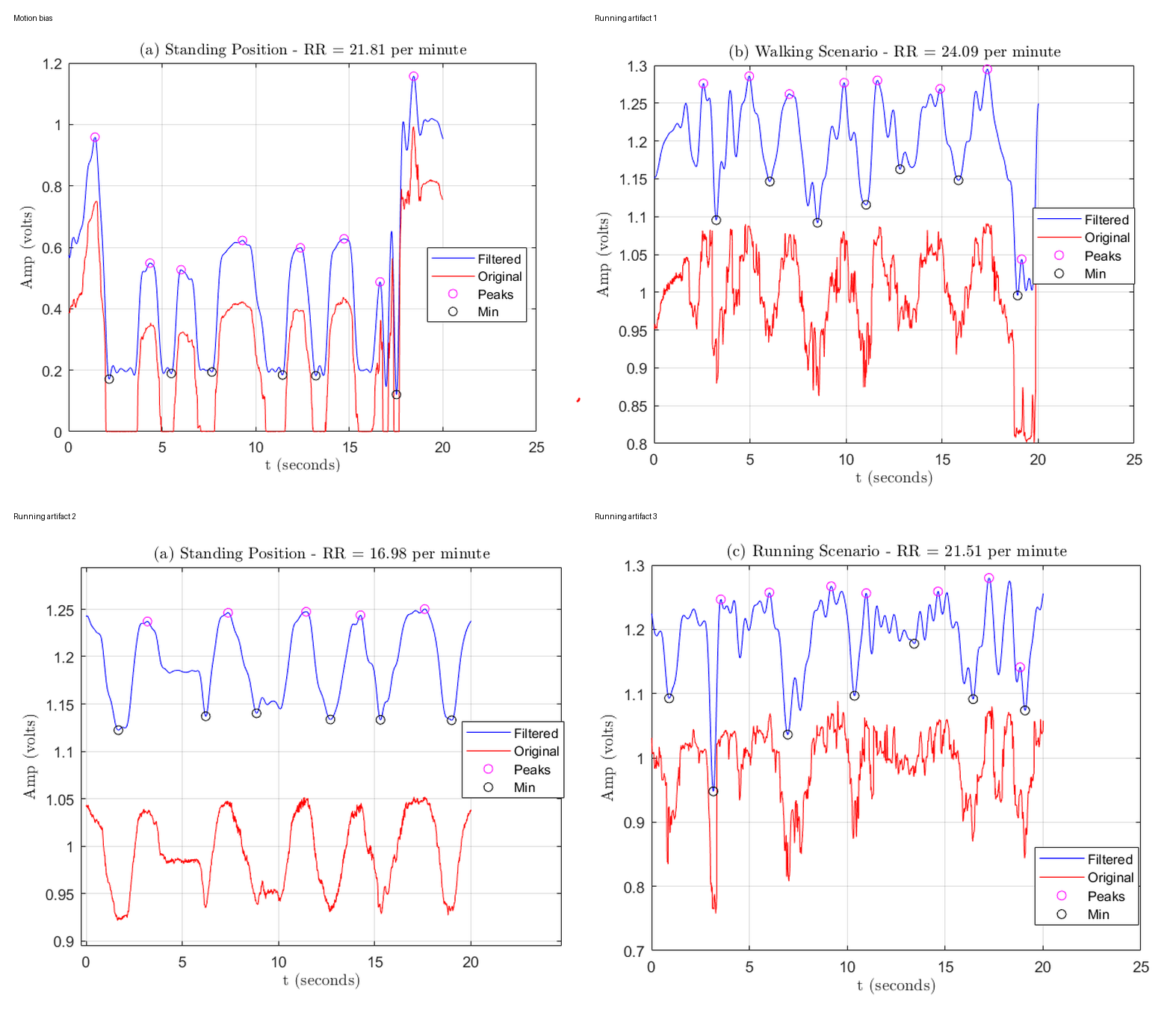}
    \caption{Examples of motion-induced baseline shifts and running artifacts from the validation experiments.}
    \label{fig:motion_artifacts}
\end{figure}

Figure~\ref{fig:motion_artifacts} shows examples where motion and placement effects become comparable to or larger than the respiratory signal. These examples provide additional context for the operating range discussed in Section~\ref{sec:posture_motion_results}.

\section{Baseline Comparison with Prior Work}
\label{app:baseline}

Table~\ref{tab:baseline_comparison} preserves the comparison reported in the original project document and corrects the first row's sensor description. Because the studies use different labels, cohorts, feature sets, and validation procedures, the values provide historical context rather than a controlled ranking.

\begin{table}[h]
    \centering
    \caption{Comparison with prior stress-detection studies. The last row corresponds to our results.}
    \label{tab:baseline_comparison}
    \resizebox{\linewidth}{!}{%
    \begin{tabular}{p{0.10\linewidth}p{0.10\linewidth}p{0.20\linewidth}p{0.24\linewidth}p{0.12\linewidth}p{0.17\linewidth}}
        \toprule
        Participants & Features & Classifier & Sensors used & Accuracy & Source \\
        \midrule
        3  & 8/8   & KNN & Heart-rate monitor & 83.3\% & \citet{choi2009heart} \\
        20 & 4/25  & Decision tree, SVM, BN & ECG, GSR, accelerometer & 92.4\% & \citet{sun2010activity} \\
        6  & 1/15  & Minimum-distance classifier & ECG & 79.9\% & \citet{boonnithi2011} \\
        40 & 1/148 & KNN, PNN & ECG, EMG, HRV, GSR, ST & 93.5\% & \citet{karthikeyan2013} \\
        46 & 9/19  & -- & RR, HRV & 74.6\% & \citet{salai2016} \\
        42 & 1/5   & LR, SVM, BN & EEG & 96.0\% & \citet{subhani2017} \\
        45 & 16/22 & ANOVA test & GSR & 89.0\% & \citet{zangroniz2017} \\
        23 & 20    & KNN, SVM & ECG, HRV & 92.75\% & \citet{sriramprakash2017} \\
        43 & 3/55  & Three-layer perceptron & Radar respiration sensor & 94.44\% & \citet{machado2018} \\
        12 & 6/28  & KNN, SVM, LDA, decision tree, one-hidden-layer NN & Wearable respiration sensor & 82.4\% & Ours \\
        \bottomrule
    \end{tabular}}
\end{table}